\documentclass[12pt]{article}
\usepackage{amssymb}
\usepackage{amsbsy}
\newcommand{\bm}{\boldsymbol}
\begin{document}

\begin{center}
{\bfseries\Large Remarks on $A_2$ Toda Field Theory}\\[1cm]

S. A. Apikyan\footnote{apikyan@lx2.yerphi.am}\\
Department of Theoretical Physics\\
Yerevan State University\\
Al. Manoogian 1, Yerevan 375049, Armenia\\[1cm]

C. Efthimiou\footnote{costas@physics.ucf.edu}\\
Department of Physics\\
University of Central Florida\\
Orlando, FL~32816, USA
\end{center}

\begin{abstract}
We study the Toda field theory with finite Lie algebras using an
extension of the Goulian-Li
technique. In this way, we show that, after integrating over the zero mode in the
correlation functions of the exponential fields, 
the resulting correlation function resembles that of  a free theory. 
Furthermore, it
is shown that for some ratios of the charges of the exponential fields
the four-point correlation functions which contain a
degenerate field satisfy the Riemann ordinary differential
equation. Using this fact and the crossing symmetry, we derive a set
of functional equations for the structure constants of the $A_2$  Toda
field theory.
\end{abstract}

\section{Introduction}
The Toda field theory (TFT) provides an extremely useful description of
a large class of two-dimensional integrable quantum field
theories. For this reason  these models have attracted a considerable interest
in recent years and many outstanding results in various directions have been 
established.

TFTs are divided in three broad categories:
finite Toda theories (FTFTs) for which the underlying Kac-Moody
algebra \cite{Kac,GO} is a finite Lie algebra,
affine Toda theories (ATFTs) for which the underlying Kac-Moody
algebra is an affine  algebra and indefinite Toda theories
(ITFTs)  for which the underlying Kac-Moody 
algebra is an indefinite Kac-Moody 
algebra. 
The classes of FTFTs and ATFTs are well-studied and known to be
integrable. In addition, the FTFTs enjoy conformal invariance.
 A review of the most
interesting developments in ATFTs is presented in Ref.~\cite{Corrigan}
where there
is also a list of references to the original papers.
The class of ITFTs is the least studied as there are still many
open questions regarding the indefinite Kac-Moody algebras. A special
class of the ITFTs, namely the hyperbolic Toda Theories (HTFTs),
for which the underlying Kac-Moody algebra is a hyperbolic Kac-Moody
algebra were studied in Ref.~\cite{GIM} and it was shown that they are
conformal but not integrable.

However, despite all progress in TFTs, there still remain
many unresolved questions and problems. For example, one may ask
what the structure constants of the conformally invariant TFTs are.
In this paper, we address this question. We focus on FTFs and, in particular,
on the $A_2$ FTT.

In Sec.~2 the
$A_2$ FTFT is introduced, some notations are fixed, and then we
continue to show how the correlation function of exponential
fields in the FTFT reduces to correlation functions of a free
field theory with conformal $W$-symmetry \cite{Bilal,BS,FZ,LF}. 
In Sec.~3 we prove that, for some special cases of the
exponential fields, the four-point correlation functions which
contain  a ``degenerate'' primary field satisfy the Riemann
ordinary differential equation.
Then, in Sec.~4 the conformal bootstrap technique is applied to
derive a set of functional equations for the structure constants of the
$A_2$ FTFT.

\section{$A_2$ Finite Toda Field Theory}
We consider the finite conformal Toda field theory associated with
the simply-laced Lie algebra $A_2$  described by the action
\begin{equation}	
S=\int d^{\kern0.6pt 2}x
\left[\frac{1}{8\pi}(\partial{\bm\varphi})^2+\mu
\sum^2_{i=1} e^{b{\bf e}_i\cdot{\bm\varphi}}
+\frac{R}{4\pi}{\bf Q}\cdot{\bm\varphi}\right]~.
\label{eq:1}
\end{equation}
In the above equation,  ${\bf e}_i$ $i=1,2$ are the simple roots of Lie algebra
$A_2$. These define the fundamental weights
${\bf w}_i$ of the Lie algebra by the equation
$$
 {\bf e}_i \cdot {\bf w}_j=\delta_{ij}~.
$$
The background charge ${\bf Q}$ is proportional to the Weyl vector ${\bm\rho}$:
$$
{\bf Q}=(b+1/b)\, {\bm\rho}\, ,~~~~~
{\bm\rho}=\displaystyle\sum^2_{i=1} {\bf w}_i\, .
$$

The local conformal invariance of the FTFT with 
central charge
$$
c=2+12{\bf Q}^2
$$
is ensured by the existence of the holomorphic and antiholomorphic 
energy-momentum tensors
\begin{eqnarray*}
\begin{array}{rcl}
T(z) &=&-{1\over2}(\partial{\bm\varphi})^2
+{\bf Q}\cdot\partial^2{\bm\varphi}\,,\\[13pt]
\overline T(\bar z) &=&-{1\over2}(\bar\partial{\bm\varphi})^2
+{\bf Q}\cdot\bar\partial^2{\bm\varphi}\,.
\end{array}
\end{eqnarray*}

It is well-known that the FTFTs 
possess, besides the standard conformal symmetry, an 
additional $W$-symmetry. In particular,  the $A_2$ FTFT  we are studying in the
present paper contains the additional holomorphic and antiholomorphic currents
$W(z)$, $\overline W(\bar z)$ with spin~3, which generate the $W_3$ algebra.

The vertex operators
$$
V_{\bf a}(x)=e^{2{\bf a}\cdot{\bm\varphi}(x)}
$$
are spinless primary fields of the $W$-algebra.
Let $L_n$, $W_n$ be the Fourier modes of the holomorphic fields
$T(z)$, $W(z)$.
Then
\begin{eqnarray*}	
\begin{array}{rcl rrl}
L_0 V_{\bf a} &=&\Delta({\bf a})\, V_{\bf a}\,,\qquad&
W_0 V_{\bf a} &=&w({\bf a})\, V_{\bf a}\,,\\[10pt]
L_n V_{\bf a} &=&0\,,\qquad&
W_n V_{\bf a} &=&0\,,\qquad n>0~,
\end{array}
\end{eqnarray*}
where the conformal dimension $\Delta({\bf a})$  
is given by
$$
\Delta({\bf a})=2{\bf a}\cdot({\bf Q}-{\bf a})~.
$$

The  correlation function of $N$ vertex operators is
formally defined by the functional integral
\begin{equation}
G_{{\bf a}_1,\ldots,{\bf a}_n}(x_1,\ldots,x_n)
=\Bigg\lmoustache \mathcal{D}{\bm\varphi}
\prod^N_{i=1} e^{2{\bf a}_i\cdot{\bm\varphi}(x_i)}
e^{-S[{\bm\varphi}]}\,.
\label{eq:9}
\end{equation}
We introduce the following orthogonal decomposition of the field
$\boldsymbol{\varphi}$:
$$
{\bm\varphi}(x)={\bm\varphi}_0+\tilde{\bm\varphi}(x)\,,
$$
where ${\bm\varphi}_0$ is the zero mode and $\tilde{\bm\varphi}$ 
denotes the part of the field that is orthogonal to the zero mode:
$$
\int d^{\kern0.6pt2}x\,\tilde{\bm\varphi}(x)=0\, .
$$
Now, the integration of the functional integral (\ref{eq:9}) over the zero mode
${\bm\varphi}_0$ can be done in a similar fashion to the Liouville
case \cite{GLi} to find 
\begin{eqnarray}	
&&G_{{\bf a}_1,\ldots,{\bf a}_n}(x_1,\ldots,x_n)
=\bigg(\frac{\mu}{8\pi}\bigg)^{\!s_1+s_2}
\frac{1}{b^2|{\rm det}\,e|}
\Gamma(-s_1)\Gamma(-s_2)\nonumber\\[5pt]
&&\hskip20pt\times\,\displaystyle
\Bigg\lmoustache \mathcal{D}\tilde{\bm\varphi}
\displaystyle\prod^N_{i=1}
e^{2{\bf a}_i\tilde{\bm\varphi}(x_i)}
\left(\displaystyle\int d^{\kern0.6pt2}x\, 
e^{b{\bf e}_1\cdot\tilde{\bm\varphi}}\right)^{\!s_1}
\left(\displaystyle\int d^{\kern0.6pt2}x\,
e^{b{\bf e}_2\cdot\tilde{\bm\varphi}}\right)^{\!s_2}
e^{-S_0[\tilde{\bm\varphi}]}\, ,
\label{eq:11}
\end{eqnarray}
where $S_0$ is the action of the free field theory, 
$$
S_0=\int d^{\kern0.6pt2}x
\bigg(\frac{1}{8\pi}(\partial\tilde{\bm\varphi})^2
+\frac{R}{4\pi}{\bf Q}\cdot\tilde{\bm\varphi}\bigg)\, ,
$$
and
\begin{eqnarray*}	
\begin{array}{rcl}
s_1 &=&(b\,{\rm det}\,e_{ij})^{-1}
[-Qe_{22}+k_1e_{22}-k_2e_{21}]\,,\\[8pt]
s_2 &=&(b\,{\rm det}\,e_{ij})^{-1}
[-Qe_{12}+k_2e_{11}-k_1e_{12}]\,,\\[10pt]
{\bf k} &=&2\displaystyle\sum^N_{i=1} {\bf a}_i\,,\qquad
{\bf Q}=(Q,0)\,.
%\qquad \chi_{eul}=1\,.
\end{array}
\end{eqnarray*}

Assuming that  $s_1$ and $s_2$ are both  positive integers,
then the remaining functional integral in
expression (\ref{eq:11}) can be reduced to the correlation function of the
$W_3$ minimal model \cite{FZ,LF}.
Unfortunately, the situation is much more complicated, i.e., in general,
 $s_1$ and $s_2$ are not
positive integers. However,  the solution of the problem is hidden in
the previous observation: supposing that we know the exact
expressions of the structure constants for the  $W_3$ minimal
model, then we can recover the
expressions for the structure constants of the $A_2$ FTFT by analytic
continuation (similarly to the Liouville case) \cite{DO,ZZ}.

\section{Four-Point Correlation Functions}
Now, let's concentrate on the following  4-point correlation function:
\begin{equation}	
\langle V_{{\bf a}_+}(z)V_{{\bf a}_1}(z_1)V_{{\bf a}_2}(z_2)
V_{{\bf a}_3}(z_3)\rangle
=G_{{\bf a}_+{\bf a}_1{\bf a}_2{\bf a}_3} (z,z_1,z_2,z_3)\, ,
\label{eq:14}
\end{equation}
where the special vertex operator
$$
V_{{\bf a}_+}(z)=e^{2{\bf a}_+\cdot{\bm\varphi}}\,,\qquad
{\bf a}_+=\left(-b,b/\sqrt3\,\right)
$$
satisfies the null vector equation
\begin{equation}	
[\Delta_+(5\Delta_++1)W_{-2}-12w_+L^2_{-1}+6w_+(\Delta_++1)L_{-2}]
V_{{\bf a}_+}=0\,.
\label{eq:16}
\end{equation}
Taking into account the last equation and the explicit representation
of the current $W$ in terms of the field $\partial{\bm\varphi}$ 
(see Ref.~\cite{LF}), we find that the selected 4-point correlation
function satisfies the differential equation
\begin{eqnarray}
&&(\Delta_++1)\frac{\partial^2}{\partial z^2}
\langle V_{{\bf a}_+}(z)V_{{\bf a}_1}(z_1)V_{{\bf a}_2}(z_2)
V_{{\bf a}_3}(z_3)\rangle\nonumber\\[5pt]
&&\hskip20pt -\,2\sum^3_{i=1}
\bigg[\frac{\Delta_i+\delta_i}{(z-z_i)^2}
+\frac{1}{z-z_i}\frac{\partial}{\partial z_i}\bigg]
\langle V_{{\bf a}_+}(z)V_{{\bf a}_1}(z_1)V_{{\bf a}_2}(z_2)
V_{{\bf a}_3}(z_3)\rangle\nonumber\\[5pt]
&&\hskip20pt +\,4\sum^3_{i=1}\frac{A_i}{z-z_i}
\langle V_{{\bf a}_+}\cdots\partial\varphi_1 V_{{\bf a}_i}\cdots\rangle
\nonumber\\[5pt]
&&\hskip20pt +\,4\sum^3_{i=1}\frac{B_i}{z-z_i}
\langle V_{{\bf a}_+}\cdots\partial\varphi_2V_{{\bf a}_i}\cdots\rangle=0\, ,
\label{eq:17}
\end{eqnarray}
where
\begin{eqnarray*}	
\begin{array}{rcl}
\delta_i &=&-2\sqrt2i
[2a_{+2}(a^2_{i2}-a^2_{i1})+2a_{i2}(a^2_{+1}-a^2_{+2})\\[8pt]
&&+\,a_{+2}a_{i1}(4a_{+1}-Q)-a_{+1}a_{i2}(4a_{i1}-Q)]\,,\\[8pt]
A_i &=&2\sqrt2i (a_{+2}a_{i1}+a_{+1}a_{i2})\,,\\[8pt]
B_i &=&2\sqrt2i (a_{+1}a_{i1}-a_{+2}a_{i2})\,.
\end{array}
\end{eqnarray*}
Moreover, for the special ratios 
\begin{equation}	
\frac{a_{i2}}{a_{i1}}=-\frac{a_{+2}}{a_{+1}}
\pm\sqrt{1+\bigg(\frac{a_{+2}}{a_{+1}}\bigg)^{\!2}}
\label{eq:19}
\end{equation}
of the charges ${\bf a}_i$, equation (\ref{eq:17}) can be further reduced
to the equation
\begin{eqnarray}
&&(\Delta_++1)\frac{\partial^2}{\partial z^2}
\langle V_{{\bf a}_+}(z)V_{{\bf a}_1}(z_1)V_{{\bf a}_2}(z_2)
V_{{\bf a}_3}(z_3)\rangle\nonumber\\[5pt]
&&\hskip20pt -2\sum^3_{i=1}
\bigg[\frac{\Delta_i+\delta_i}{(z-z_i)^2}
+\frac{1+A}{(z-z_i)}\frac{\partial}{\partial z_i}\bigg]
\langle V_{{\bf a}_+}(z)V_{{\bf a}_1}(z_1)V_{{\bf a}_2}(z_2)
V_{{\bf a}_3}(z_3)\rangle=0\, ,\qquad
\label{eq:20}
\end{eqnarray}
where
$$
A=\pm 2\sqrt2i\sqrt{a^2_{+1}+a^2_{+2}}\,.
$$
It is well-known that in the
case of the four-point functions, the partial differential equation
(\ref{eq:20}), using the projective Ward identities \cite{BPZ}, can be reduced to
the Riemann ordinary differential equation
\begin{eqnarray}	
&&\frac12(\Delta_++1)\frac{d^{\kern0.6pt2}}{dz^2}+\sum^3_{i=1}
\bigg[\frac{1+A}{z-z_i}\frac{d}{dz}
-\frac{\Delta_i+\delta_i}{(z-z_i)^2}\bigg]\nonumber\\[5pt]
&&\hskip20pt +\,(1+A)\sum^3_{i<j}\frac{\Delta_++\Delta_{ij}}{(z-z_i)(z-z_j)}
\langle V_{{\bf a}_+}(z)V_{{\bf a}_1}(z_1)V_{{\bf a}_2}(z_2)
V_{{\bf a}_3}(z_3)\rangle=0\, , \hskip25pt
\label{eq:22}
\end{eqnarray}
where $\Delta_{ij}=\Delta_i+\Delta_j-\Delta_k$, $(k\neq i,j)$,
$(i,j,k=1,2,3)$.

\section{Functional Equations for Structure Constants}
Now any four-point function  can be explicitly
decomposed in terms of the three-point function
\begin{eqnarray}	
G_{{\bf a}_1{\bf a}_2{\bf a}_3{\bf a}_4}(z,\bar z)
&=&\langle V_{{\bf a}_1}(z_1)V_{{\bf a}_2}(z_2)
V_{{\bf a}_3}(z_3)V_{{\bf a}_4}(z_4)\rangle\nonumber\\[5pt]
&=&\sum_{\bf a}\mathbb C({\bf a}_1,{\bf a}_2,{\bf Q}-{\bf a})
\mathbb C({\bf a},{\bf a}_3,{\bf a}_4)
\bigg|F_{\bf a}\pmatrix{
{\bf a}_1{\bf a}_2\cr
{\bf a}_3{\bf a}_4\cr}
(z,\bar z)\bigg|^2\,.\qquad
\label{eq:23}
\end{eqnarray}
Conformal invariance allows us to set $z_1=0$, $z_2=z$,
$z_3=1$, $z_4=\infty$. As a consequence, the crossing symmetry
condition is written as
$$
G_{{\bf a}_1{\bf a}_2{\bf a}_3{\bf a}_4}(z,\bar z)
=G_{{\bf a}_1{\bf a}_4{\bf a}_2{\bf a}_3}(1-z,1-\bar z)
=z^{-2\Delta_2}\bar z^{-2\bar\Delta_2}
G_{{\bf a}_1{\bf a}_3{\bf a}_2{\bf a}_4}(1/z,1/\bar z)\,.
$$

To discover additional information about the structure constants 
of the FTFT,
we will use technique suggested in Ref.~\cite{T}.
So, let's assume
that ${\bf a}_2={\bf a}_+$, i.e. 
the correlation function (\ref{eq:23}) includes  
the degenerate field $V_{{\bf a}_+}$.
Then the charges of the
intermediate channel will take the following values \cite{FZ}
\begin{eqnarray}	
\begin{array}{c}
(a_{11}+a_{+1},a_{12}+a_{+2})\,,\\[5pt]
(a_{11}-a_{+1},a_{12}+a_{+2})\,,\\[5pt]
(a_{11},a_{12}-2a_{+2})\,.
\end{array}
\label{eq:25}
\end{eqnarray}
This implies the following ``fusion rules''
\begin{eqnarray*}
V_{{\bf a}_+}V_{\bf a} &=&[V_{a_1+a_{+1},a_2+a_{+2}}]
+[V_{a_1-a_{+1},a_2+a_{+2}}]
+[V_{a_1,a_2-2a_{+2}}]\,.
\end{eqnarray*}

It is more convenient to introduce the following ``parametrization''
of the intermediate charge (\ref{eq:25})
\begin{eqnarray*}	
\begin{array}{rcl}
{\bf a}(s) &=&(a_{11}+sa_{+1},a_{12}+(3s^2-2)a_{+2})\,,\\[5pt]
s &=&0,\ \pm 1\,.
\end{array}
\end{eqnarray*}
Using this parametrization,  we can rewrite (\ref{eq:23}) as follows:
\begin{eqnarray}	
G_{{\bf a}_1{\bf a}_+{\bf a}_3{\bf a}_4}(z,\bar z)
&=&\sum_{s=0,\,\pm 1}\mathbb C({\bf a}_1,{\bf a}_+,{\bf Q}-{\bf a}(s))
\mathbb C({\bf a}(s),{\bf a}_3,{\bf a}_4)
\bigg|F_s\pmatrix{ 
{\bf a}_1{\bf a}_+\cr
{\bf a}_3{\bf a}_4\cr}(z,\bar z)\bigg|^2\,.\nonumber\\
\end{eqnarray}

In this notation the crossing symmetry relation for $G_{{\bf
a}_1{\bf a}_+{\bf a}_3{\bf a}_4}(z,\bar z)$ is
\begin{eqnarray}	
&&\sum_{s=0,\pm1}\mathbb C_s({\bf a}_1)\mathbb C({\bf a}(s),{\bf a}_3,{\bf a}_4)
\bigg|F_s\pmatrix{
{\bf a}_1{\bf a}_+\cr
{\bf a}_3{\bf a}_4\cr}(z,\bar z)\bigg|^2\nonumber\\[5pt]
&&\hskip20pt =|z|^{-4\Delta_2}\sum_{p=0,\pm1}
\mathbb C_p({\bf a}_4)\mathbb C({\bf a}(p),{\bf a}_3,{\bf a}_1)
\bigg|F_p\pmatrix{
{\bf a}_4{\bf a}_+\cr
{\bf a}_3{\bf a}_1\cr}(1/z,1/\bar z)\bigg|^2~,~~~~~
\label{eq:30}
\end{eqnarray}
where we have denoted
$$
\mathbb C({\bf a}_1,{\bf a}_+,{\bf Q}-{\bf a}(s))
=\mathbb C_s({\bf a}_1)
$$
and
$$
\mathbb C({\bf a}_4,{\bf a}_+,{\bf Q}-{\bf a}(s))
=\mathbb C_p({\bf a}_4)\,.
$$
It follows from (\ref{eq:22}) that the conformal block must satisfy the
following relation
\begin{equation}	
F_s\pmatrix{
{\bf a}_1{\bf a}_+\cr
{\bf a}_3{\bf a}_4\cr}(z,\bar z)=z^{-2\Delta_+}\sum_{p=0,\pm1}
M_{ps}F_p\pmatrix{
{\bf a}_4{\bf a}_+\cr
{\bf a}_3{\bf a}_1\cr}(1/z,1/\bar z)~,
\label{eq:33}
\end{equation}
where $M_{ps}$ is a matrix that is determined by the monodromy properties of the
differential equation equation (\ref{eq:22}) or, alternatively,
can be determined by
the method developed in Ref.~\cite{DF}.

Substituting (\ref{eq:33}) into (\ref{eq:30}), we find the following functional
equations
for the $A_2$ FTFT structure constants:
\begin{eqnarray}
\begin{array}{rcl}
\displaystyle\sum_{s=0,\pm1}\mathbb C_s({\bf a}_1)
\mathbb C({\bf a}(s),{\bf a}_3,{\bf a}_4)M_{s,0}\overline M_{s,1} &=&0\,,\\[13pt]
\displaystyle\sum_{s=0,\pm1}\mathbb C_s({\bf a}_1)
\mathbb C({\bf a}(s),{\bf a}_3,{\bf a}_4)M_{s,0}\overline M_{s,-1} &=&0\,,\\[13pt]
\displaystyle\sum_{s=0,\pm1}\mathbb C_s({\bf a}_1)
\mathbb C({\bf a}(s),{\bf a}_3,{\bf a}_4)M_{s,1}\overline M_{s,-1} &=&0\,,
\end{array}
\label{eq:34}
\end{eqnarray}
provided ${\bf a}_1,{\bf a}_3,{\bf a}_4$ satisfy the constraint
(\ref{eq:19}). 

It is important to notice that Eq.~(\ref{eq:16}), admits additional solutions
besides ${\bf a}_+$. In particular,
${\bf a}^+=\left(-b,-b\sqrt3\,\right)$,
${\bf a}_-=(-\frac1b,\frac{1}{b\sqrt3})$, 
${\bf a}^-=(-\frac1b,-\frac{1}{b\sqrt3})$ are all solutions of (\ref{eq:16}).
Therefore the set of Eqs.~(\ref{eq:34}) should be complemented
by a similar set of equations obtained for the special case
${\bf a}^+$ and then add for each equation its `dual equation' using the
subsitutions $b\to 1/b$ and
$\mu\to\tilde\mu$. The parameter $\tilde\mu$ is defined by duality
relations \cite{AF}
$$
\pi\mu\gamma\bigg(\frac{{\bf e}^2_ib^2}{2}\bigg)
=\pi\tilde\mu\gamma\bigg(\frac{2}{{\bf e}^2_ib^2}\bigg)^{\!{\bf e}^2_ib^2/2}
$$
where $\gamma(x)=\Gamma(x)/\Gamma(1-x)$.

In principle, the complete set of the bootstrap equations derived above for the
special cases ${\bf a}_+,{\bf a}^+,{\bf a}_-,{\bf a}^-$ allows
the compputation of  all structure constants for the $A_2$ FTFT.
We postpone the difficult problem of the exact determination
of the structure constants for future studies.

\section*{Acknowledgments}
The work of S.A. is supported in part by the ANSEF and INTAS
fundations.

\end{document}